# How Tuning Interfaces Impacts Dynamics and Structure of Polymer Nanocomposites Simultaneously

Anne-Caroline Genix,[1,*] Vera Bocharova,[2] Bobby Carroll,[2] Philippe Dieudonné-George,[1] Edouard Chauveau,[1] Alexei P. Sokolov,[2,3] and Julian Oberdisse [1]

[1]Laboratoire Charles Coulomb (L2C), Université de Montpellier, CNRS, F-34095 Montpellier, France

[2]Chemical Sciences Division, Oak Ridge National Laboratory, Oak Ridge, TN 37831, USA

[3]Department of Chemistry, University of Tennessee, Knoxville, TN 37996, USA

* Corresponding author: anne-caroline.genix@umontpellier.fr

**Abstract**

Fundamental understanding of macroscopic properties of polymer nanocomposites (PNCs) remains difficult due to the complex interplay of microscopic dynamics and structure, namely interfacial layer relaxations and three-dimensional nanoparticle arrangements. The effect of surface modification by alkyl methoxysilanes at different grafting densities has been studied in PNCs made of poly(2-vinylpyridine) and spherical 20 nm silica nanoparticles (NPs). The segmental dynamics has been probed by broadband dielectric spectroscopy, and the filler structure by small-angle X-ray scattering and reverse Monte Carlo simulations. By combining the particle configurations with the interfacial layer properties, it is shown how surface modification tunes the attractive polymer-particle interactions: bare NPs slow down the polymer interfacial layer dynamics over a thickness of ca. 5 nm, while grafting screens these interactions. Our analysis of interparticle spacing and segmental dynamics provides unprecedented insight into the effect of surface modification on the main characteristics of PNCs: particle interactions and polymer interfacial layers.

## 1. INTRODUCTION

The physical properties of polymer nanocomposites (PNCs) depend on those of the particles, the polymer, and their interplay. [1-4] The latter may be rationalized by the structural and dynamical properties of the polymer interfacial layer, which may have, e.g., a density, [5, 6] a mechanical modulus [7-9] or a glass-transition temperature ($T_g$) [10, 11] different from the neat polymer. [12] Chemical modification of the surface of the nanoparticles (NPs) by grafting of small molecules have been performed in many different experimental systems, typically as coating agents for improvement of nanoparticle dispersion in industrial PNCs for tire applications, [13] or in order to study this effect in model systems. [14, 15] The main effect of surface modification is to change the polymer-NP interactions by reactions with surface groups like silanols on silica, or adding a layer of different dielectric constant, [16] or hydrophobicity. [17] In the latter case, the aim of the grafting is to compatibilize inherently hydrophilic NPs with the polymer, allowing for a more intimate mixture, i.e., better dispersion. However, some PNCs have naturally strong NP-polymer interactions, namely due to hydrogen bonding. In this case, the opposite effect of surface modification on the dispersion may be expected, with a possible impact on polymer dynamics. It is the aim of the present paper to investigate this effect by a quantitative combination of two experimental methods, one for the nanoparticle structure, and the other for the polymer dynamics, analyzed with dedicated quantitative numerical tools.

There is a large literature on the dynamics of polymer nanocomposites, making use of a variety of techniques. Macroscopic mechanical properties are commonly used to characterize the modulus, flow properties and the segmental relaxation ($\alpha$) associated with the glass transition of the polymer. [18, 19] Neutron scattering techniques like backscattering [20, 21] or spin-echo spectroscopy [22, 23] have the great advantage of combining spatial with dynamical information. On the other hand, broadband dielectric spectroscopy (BDS) is a laboratory technique, which gives direct access to dipole relaxation in the samples, as well as ionic conductivity, in a broad frequency range (typically $10^{-2} – 10^7$ Hz). [24] It has been abundantly used to investigate the segmental dynamics in PNCs. [7, 25-29] In PNCs of interest for the present study, there is a succession of dynamical processes of different origin at a given temperature. At low frequencies, Maxwell-Wagner-Sillars (MWS) processes are related to polarization of the NP interfaces and charge transport along them [24, 30]. Then the α-relaxation of the polymer – with or without an interfacial component – is found; lastly, the β-relaxation at the highest frequencies is the consequence of local molecular motions. In a recent contribution, we have evidenced the existence of two qualitatively different MWS processes in PNCs, which allowed a dynamical view of NP percolation through the samples. [31, 32] These dynamical processes are conveniently described by the Havriliak-Negami formalism. [33] The same is true for the symmetric β-process, which however has not always been included in the total description of the dielectric response. [7] The α-process observed in PNCs, finally, is usually affected by the presence of nanoparticles with a significant broadening on the low-frequency side (longer timescales) when NP-polymer interactions are attractive and favor physical adsorption of the chains. [12] The interlayer model (ILM) proposed by Steeman et al [34] has been applied successfully to heterogenous polymer dynamics in PNCs. [35] This non-additive three-phase model describes the polymer α-relaxation using two contributions, one for the bulk, and one for the (slower) interfacial layer, together with the filler contribution. Any underlying spatial gradient of the relaxation time close to the particle surface is thus represented by a box-like profile. The ILM approach provides the relaxation time distribution as well as the volume fraction of the interfacial layer from which one can deduce the layer thickness, typically in the range 1 – 5 nm, assuming a given geometry, e.g., simple cubic or random NP arrangements. [7, 36] Here, we propose accessing the full NP configuration through numerical simulation-based analysis of the experimental scattering. This allows for a more realistic determination of the thickness by (numerically) coating progressively each particle of known position until reaching the targeted layer fraction.

The dynamical properties of the interfacial layer in PNCs with attractive interactions are now well characterized, but the underlying mechanism linked to the strength of the NP-polymer interaction is still a matter of discussion. Van der Waals forces can be estimated (see SI), but also the simple presence of hard (or soft) non-interacting interfaces, possibly with surface roughness, has been shown by molecular theories to impact the monomeric cage constraints and the collective elastic barrier, and thus the segmental dynamics close to the interface. [37-39] A commonly accepted description is a strong – double-exponential – spatial gradient of the relaxation time, which represents the transfer and weakening of modified cage motion from one layer to the next, typically over some 5 to 10 monomeric diameters. In the experimental ILM approach, such gradients are averaged over a given nanometric thickness corresponding to the same spatial extension. Some experimental studies have explored the interfacial properties of PNCs containing nanoparticles with grafted chains, i.e., the strongest interaction case, revealing a complex behavior and a dependence on molecular weight of the grafted chains. [40-42] Recent works by Roth et al [43, 44] present a mapping of the local $T_g$ as a function of the distance from a solid surface with grafted polymer chains, using fluorescent dyes. The authors show that there is a significant slowing-down at the interface (ca. 45 K increase in local $T_g$) which extends over an unexpectedly large length scale of about 100 nm before recovering the bulk property. [43] They propose that the interfacial perturbations relate to chain connectivity across the interface altering the energy barrier for cooperative rearrangements. On the contrary, the impact of NP surface modification with small molecules on the dynamics and spatial extent of the interfacial layer in attractive PNCs has been rarely considered, or not compared to the bare particles. [45] This is the purpose of our study, which also follows the filler structure simultaneously. Note that coating of the silica NPs implies the presence of a surface layer of different physical chemistry in terms of dielectric strength, hydrophobicity, rugosity and hardness (which can be interpreted as affecting dynamical caging constraints [37, 38]), … In the following, we therefore consider the screening effect of the grafted layer as a collection of these different effects, which cannot be disentangled in our experiments.

Small-angle scattering is one of the most powerful techniques to analyze PNC structures. Via the natural scattering contrast between NPs and polymer, it is possible to use small-angle X-rays scattering (SAXS) to characterize NP dispersions. [46] By adapting the scattering contrast chemically in small-angle neutron scattering, it is possible to either focus on the NP dispersion, or to highlight polymer conformations. [47] In the presence of crowded NP environments, which are precisely those of interest for applications, the quantitative analysis of the scattered intensity is conceptually difficult. This is due to the presence of interactions between usually polydisperse NPs, which depend among others on NP-polymer interactions. The strength of the interfacial interaction thus influences the spatial organization of the particles, and three main polymer-mediated states have been predicted by the polymer reference interaction site model (PRISM): aggregated NPs with NP-NP contact, dispersed NPs sterically stabilized by adsorbed polymer layers, and polymer-bridged NPs. [48] High interfacial attraction leads to network formation and possibly phase separation, in agreement with the observation of a low-q upturn at the highest loading by SAXS measurements. [49] Traditional SAXS data analysis is often based on reading off the position $q_0$ of a potentially weak and broad structure factor peak characteristic of nearest neighbor NP interactions, ignoring any low-q upturns indicating imperfect dispersions. $2\pi/q_0$ then gives the typical center-to-center distance, which can be converted into the interparticle surface-to-surface spacing, IPS = $2\pi/q_0 - 2R_{NP}$. Here $R_{NP}$ is the monodisperse particle radius, monodispersity being usually an approximation. Moreover, this structure factor can only be apparent – it is obtained by division of the measured intensity by the average particle form factor. It thus does not account properly for polydispersity (see SI). Last but not least, the particle form factor itself is not always well-known, as it is not systematically measured in, e.g., molecular solvents, and it can moreover be affected by the existence of polymer density variations at the NP surface. [6] While the latter point is difficult to circumvent, it is possible to work with a measurement of the polydisperse form factor, and weight all correlations correspondingly. By fitting the measured intensity using a reverse Monte Carlo (RMC) method, [50, 51] it is then possible to not only describe the possibly broad peak, but also the correlation hole, [52] and the low-q upturn, if present. Representative sets of particle positions compatible with the entire scattered intensity curve are thus obtained. A direct evaluation of the scattered intensity by simulation methods like the RMC procedure and analysis proposed here is rather unique in PNCs. Another approach is based on PRISM modeling predicting the whole set of partial structure factors which may be quantitatively confronted to experimental data. [53] In the present paper, an original development for the understanding of interfacial layers is proposed: the NP positions are analyzed in terms of nearest neighbor correlations, i.e., IPS distribution functions are calculated. An alternative approach was applied recently based on the calculation of the pore size distribution in NP/polymer mixtures with attractive interactions. [54] A yet different way of characterizing how close any monomer is to the next filler particle surface has been used by Schneider et al. [55] Both the latter and pore analysis highlights the existence and size of filler-free polymer domains, whereas the IPS distribution provides information on how close particle are.

In this article, the polymer dynamics in PNCs made of poly(2-vinylpyridine) (P2VP) and silica nanoparticles of various degrees of surface modification by alkyl methoxysilanes is first studied by BDS. In the second part, measurements by SAXS of the interparticle spacing distribution functions are presented and analyzed by a reverse Monte Carlo algorithm, which allowed to combine the BDS with the SAXS results, giving access to the true thickness of the polymer interlayer.

## 2. RESULTS AND DISCUSSION

Three sets of nanocomposites with various silica volume fractions (15, 20 and 30%v) were prepared by solvent casting using bare or surface-modified NPs. All samples are listed in Table 1 with the experimentally determined silica content and the grafting densities of silane molecules ($C_{18}$ or $C_8$). The segmental relaxation of these samples has been investigated by BDS, using two different approaches for data analysis. First, by applying a model of homogeneous polymer dynamics, the average relaxation time of the polymer is followed. This description is then further refined with ILM, which takes the dynamical modification of the interfacial layer only into account. As a result, its thickness and segmental relaxation time are obtained. The NP dispersions were then studied by SAXS and RMC, and analyzed in terms of the interparticle spacings.

**Table 1.** NP volume fractions and grafting densities determined by TGA.

|  | Bare | $C_{18}$ 0.5/nm$^2$ | $C_{18}$ 1.1/nm$^2$ | $C_8$ 1.7/nm$^2$ |
|---|---|---|---|---|
| **15%v-series** | 13.5% | 13.7% | 16.3% |  |
| **20%v-series** | 19.6% | 20.4% | 18.8% |  |
| **30%v-series** | 29.2% | 28.2% | 25.0% | 26.8% |

**2.1 Dielectric spectroscopy analysis.** BDS was employed to follow the relaxations of neat P2VP and P2VP PNCs in a wide temperature range from 433 K to 233 K. In neat P2VP, two dielectric processes are observed in this range: the $\alpha$ relaxation at high temperature and the secondary ($\beta$) relaxation at low temperature (i.e., higher frequency). The latter is attributed to rotation of the pendant pyridine group of P2VP. [56] Due to the high dipole moment of this group, [57] the secondary process displays a significant dielectric strength ($\Delta\varepsilon_\beta \sim 2.2$ at room temperature), which has to be taken into account at high temperature when overlapping with the segmental relaxation. Therefore, the real and imaginary parts of the permittivity of neat P2VP were fitted simultaneously by the sum of two Havriliak–Negami (HN) functions for the α and β processes and a purely dissipative d.c. conductivity term leading to a $\omega^{-1}$ dependence [24]

$$\varepsilon^*(\omega) = \varepsilon_\infty + \sum_{i=\alpha,\beta} \frac{\Delta\varepsilon_i}{\left[1+\left(i\omega\tau_{HN_i}\right)^{\gamma_i}\right]^{\delta_i}} - i\frac{\sigma_{dc}}{\varepsilon_0\omega} \quad (1)$$

$\tau_{HN}$ is the characteristic time and γ and δ are the width and asymmetry parameters of the HN distribution, respectively. In the following, the relaxation times are defined by $\tau_{HN}$ related to the peak position in frequency $f_{max}$, which is used to determine the relaxation time $\tau_{max} = 1/(2\pi f_{max})$. The β-process was described individually in the low-T range (303 – 233 K), where it can be observed alone in the accessible frequency window, using a single symmetrical HN function also called Cole-Cole function ($\delta_\beta$ = 1, see fits in SI). A width parameter of ca. 0.2 was found corresponding to a broad time distribution, which gets slightly broader when decreasing the temperature, whereas the dielectric strength was found to barely vary. The temperature dependence of $\tau_\beta$ follows an Arrhenius behavior with an activation energy of 56 kJ/mol and a prefactor of $2\times10^{-15}$ s, in good agreement with previous studies. [58-60]

In nanocomposites, a third symmetrical HN-process was systematically added in eq 1 to describe the interfacial Maxwell–Wagner–Sillars (MWS) process located at frequencies much below the α-relaxation and usually covered by conductivity. This process is associated with polarization effects in the presence of particles. [61] The β-process in PNCs was characterized in the same way as in pure P2VP. It occurs nearly at the same frequency but with a small, systematic shift towards higher frequency and a constant activation energy (see SI). A lower intensity – consistent with the lower polymer fraction – and a slightly broader time distribution ($\gamma_\beta$ = 0.18 ± 0.01) were obtained in PNCs. These features were observed independently of both the silica content and grafting density. One may note that an enhancement of the local dynamics (picosecond and β-relaxations) was reported in a similar PNC with bare NPs whereas, at the same time, an interfacial layer of slowed-down segmental dynamics was observed. [60] It means that the polymer dynamics on a variety of time scales is affected differently in presence of silica, possibly due to a complex interplay between density effects and chain stretching. These aspects are however outside the scope of the present study, where our intention is to focus on the segmental dynamics. In all cases (neat and PNCs), the distribution of relaxation times associated with the β-process was extrapolated from low-T to the higher temperature range, where the segmental relaxation is well-centered in the frequency window (368 – 433 K). This was done to reduce the number of fitting parameters in eq 1 while accounting for the high-frequency contribution of the secondary dynamics to the α-process, which quantitatively affects its dielectric force.

The description of the segmental relaxation in PNCs is the central part of our analysis. In a first approach, it can be described by a single HN-contribution in eq 1. Assuming that polymer dynamics is not affected by the NPs and the attractive NP-polymer interactions, i.e., describing it by the neat polymer as illustrated in Scheme 1a, immediately contradicts the data. [12] To fit experimental results, it is necessary to introduce a polymer with dynamics different from the one of the neat polymer, as shown in Scheme 1b. In this representation, the entire polymer part in PNC is thought to display a homogeneously modified response with respect to the pure matrix, and a single time distribution is obtained for the segmental relaxation. In Scheme 1c, finally, the contributions from two phases are considered: the interfacial layer and the unmodified bulk polymer far from the particles. In this case, the contributions of each component are not additive, and the interference terms are explicitly taken into account in the interfacial layer model for heterogeneous systems.[34] The detailed equations of this model are given in [35] (see also in SI). For the dielectric response of the bulk polymer in PNC, we used the HN-function of the neat polymer (all HN-parameters fixed). Then, the free parameters in ILM are the volume fraction of interfacial layer, $\Phi_{IL}$, and the dielectric function of the interphase, $\varepsilon_{IL}^*(\omega)$, which is well-described by a symmetrical HN-process.[7] Note that the dielectric strength and the quantity of the interfacial layer have different effects on the dielectric response. While an increase in the dielectric strength in the interfacial layer leads only to an increase of the interfacial layer contribution in the intermediate frequency range, an increase in the volume fraction of interfacial layer is concomitant with a decrease of the bulk polymer fraction also modifying the high-frequency side of the α-process. The β-process was also included in ILM using an extrapolation of the low-T data of each sample as mentioned above.

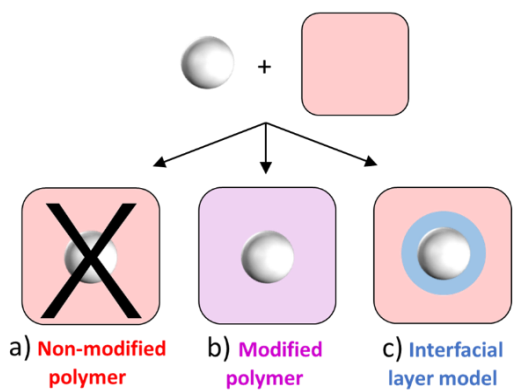

**Scheme 1.** Schematic representation of the different approaches to describe the effect of the NPs on the polymer segmental relaxation. **a)** Unmodified polymer dynamics. **b)** Homogeneously modified polymer dynamics. **c)** Locally modified polymer dynamics. Schemes b and c are used to analyze the dielectric spectra of P2VP-silica PNCs.

**2.2 Scattering analysis: RMC and IPS.** The experimental SAXS intensity I(q) was described by fully taking polydispersity into account.[62] It is convenient to plot and analyze the apparent structure factor S(q), which is obtained by dividing I(q) by the average form factor P(q). Due to the division of two small numbers at high q, S(q) tends to be noisier in this range. The relevant information on particle dispersion, however, is in the low-q domain, up to the nearest neighbor interaction peak. All the data are described using a reverse Monte Carlo simulation, following a development of previous approaches.[50, 51] In short, N spherical particles obeying the experimental log-normal size distribution are placed in a cubic simulation box with periodic boundary conditions, of dimension $L_{box} = 2\pi/q_{min}$, where $q_{min}$ is the experimental minimum q-value, such that the total volume fraction $\Phi_{NP}$ corresponds to the experimental one of the sample. The box size thus matches the information provided by the experimental data. The scattered intensity of the particles in the simulation box is calculated using a combination of the Debye formula [63] at high q, and a lattice calculation avoiding box contributions [64-66] at low q. Particles are then moved around randomly while following a simulated annealing procedure searching for the best agreement of the theoretically predicted apparent structure factor with the experimental one. This optimization followed by equilibration is quantified by the evolution of $\chi^2$.[64] Note that there are no interaction potentials other than the respect of hard-core repulsion between particles, and the simulation is only guided by the search for agreement with the experimental scattering. In the end, S(q) and any statistical measures like particle distances are averaged over different particle configurations and represent the result of the simulation.

For a monodisperse assembly of particles, one would naturally study the pair-correlation function, i.e., the distribution function of the centers-of-mass. For comparison, the pair-correlation functions of our polydisperse systems are reported in the SI. Note that they are smoothed by polydispersity, and that they do not represent the Fourier transform of the apparent structure factor. As we are interested in the space available to the polymer between polydisperse nanoparticles, and as this space depends both on the center-to-center distance, and on the respective particle sizes, we have directly determined the surface-to-surface distribution function, which we call IPS, for interparticle spacing. This function is determined by binning and averaging over different configurations. The first bin is from zero (particles in touch) to 0.1 nm of interparticle distance (surface-to-surface), and all further bins are exponentially increasing in thickness (x 1.05 each time). Direct contact means that surfaces are separated by no more than 0.1 nm. Note that this uncertainty is in line with the overall uncertainty of SAXS measurements given by the experimental q-spacing, maximum values, and error bars. Such a determination of IPS is a generalization of the standard calculation of well-known structures (cubic, random…) with a power law in volume fraction, and a dependence on the maximum packing fraction of particles.

**2.3 Average polymer dynamics in PNCs.** The dynamical response of the polymer part of the nanocomposites was investigated by dielectric spectroscopy as a function of frequency and temperature. In Figure 1, the experimental dielectric loss functions at 423 K are shown for the three silica concentrations, 15, 20, and 30%v, respectively, and various surface-modifications by $C_{18}$-silane, from bare to 1.1 $nm^{-2}$. At this temperature, all the relevant processes are located in the experimental frequency window. Dielectric spectra at other temperatures are shown in the SI, and results are summarized in this article in the form of a relaxation map including all temperatures.

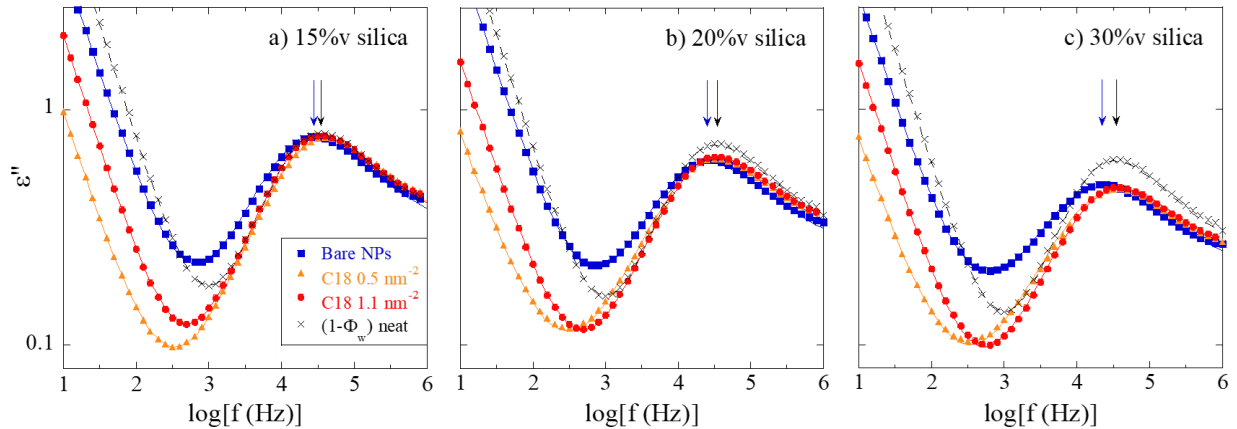

**Figure 1.** Comparison of the dielectric loss spectra of neat P2VP (black crosses, data normalized to the weight polymer fraction, 1 - $\Phi_w$) and PNCs with different surface modifications of the silica NPs for the series with $\Phi_{NP}$ = 15 **(a)**, 20 **(b)** and 30%v **(c)** at 423 K. The black dashed line is a fit with eq 1 ($\alpha$, $\beta$ and conductivity), while the solid lines represent the fit for homogeneous polymer dynamics (Scheme 1b) with the sum of 3 HN-functions for MWS, $\alpha$ and $\beta$-processes (the latter being outside the frequency window at this temperature) and a conductivity term. Arrows indicate the maximum of the loss peak in neat polymer (black) and PNCs with bare NPs (blue).

The first characteristics of the dielectric loss functions shown in Figure 1 are that there is a visible impact of the silica content on the main relaxation peak associated with the segmental dynamics. As reported in previous works, [7, 12] its position is seen to move to lower frequencies indicating a slow-down effect with respect to the neat polymer upon introduction of the bare particles. The shift of this peak is stronger at higher silica fraction as highlighted by the arrows in Figure 1. In parallel, its intensity is reduced with a significant broadening on the low-frequency side. Secondly, a striking point is that the peak then moves back to higher frequency in PNCs with grafted NPs, i.e., surface modification induces re-acceleration of the polymer dynamics. This is accompanied by a strong reduction of the peak broadening for the surface-modified samples, which are also quite similar.

As discussed in the introduction, the shape of the curves in Figure 1 can be understood as a superposition of different processes: low frequency contributions are caused by ionic conductivity and interfacial polarization (MWS), the well-defined peak at higher frequency is due to the α-relaxation of the polymer, and its right-hand shoulder comes from the secondary β-relaxation. In Figure 1, the α-peak is modified in presence of silica, and it cannot be described by adding (see Scheme 1a for illustration) the low-frequency MWS process to the neat polymer relaxation. There is another contribution at intermediate frequency representing the polymer interfacial response, which is thus slower than the pure polymer one. From our data in Figure 1, it can be immediately concluded that surface grafting affects this interfacial response by compensating for the slow-down induced by the bare NPs. The first way of accounting for these modifications of the interfacial dielectric response is to treat the polymer as a homogenously slowed-down matrix (Scheme 1b). The second way based on a heterogeneous system (Scheme 1c), with an unmodified polymer component far from the particles and an interfacial polymer layer of different dynamics, will be presented in the next subsection. The effect of both approaches is to describe the experimental minimum in dielectric loss seen in Figure 1, and to account for the shift in the α-relaxation.

In order to capture the modified average polymer dynamics in PNCs (Scheme 1b), a simple model with an additional HN function in eq 1 for the MWS contribution is applied to the data. The results are very good fits as shown by the superposition in Figure 1 to each data set. Details of the fits are displayed in Figure 2a, where the different contributions – MWS, α, β, and conductivity – to the dielectric loss are highlighted for bare NPs (top), and grafted ones (bottom, 1.1 nm$^{-2}$) in 20%v-PNCs at 410 K. The simultaneous good fit of the real part of the dielectric permittivity ε' is also shown in each case. Fits of the dielectric spectra of 15 and 30%v-PNCs at the same temperature are given in SI.

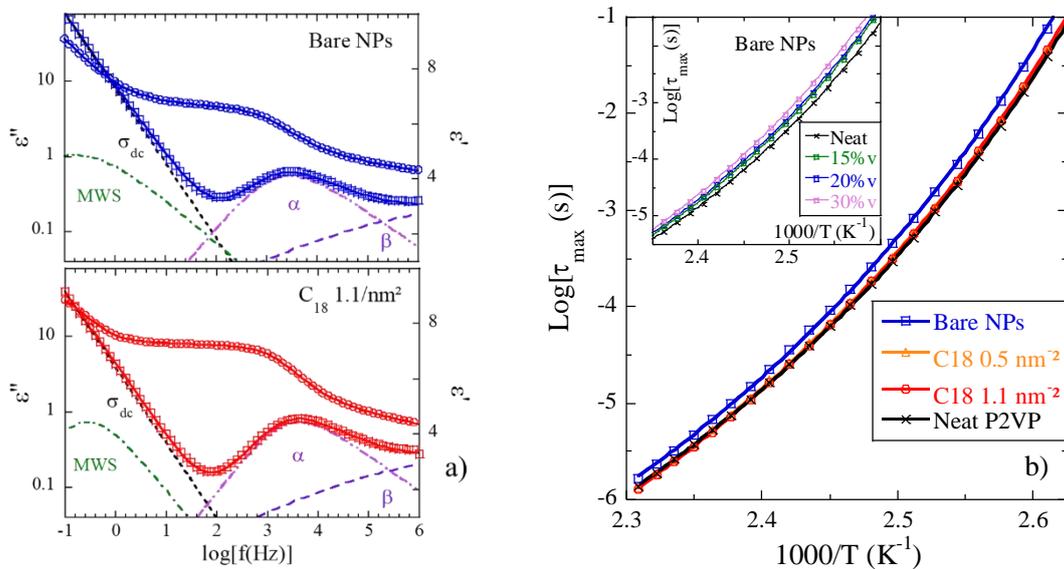

**Figure 2. a)** Frequency dependence of the real (circles, right axis) and imaginary (squares, left axis) part of the dielectric permittivity in 20%v-PNCs (T = 410 K, bare NPs top part, C$_{18}$ 1.1/nm² grafted NPs bottom part). The solid lines represent the simultaneous fits of ε' and ε" based on eq 1 using three HN processes (dashed lines for MWS, α- and β-processes as indicated) and a d.c. conductivity term (dotted line). **b)** Temperature dependence of the segmental relaxation times of 20%v-PNCs with their fits by the VFT equation. Inset: same representation for the relaxation times of PNCs with bare NPs at various volume fractions.

The parameters of these fits are reported in Table S3 in the SI for the α-process. Both shape parameters are modified in PNCs with the time distribution that gets wider and less asymmetric with respect to the neat polymer for the PNC with bare NPs, whereas grafting has the opposite effect of bringing back $\gamma_\alpha$ and $\delta_\alpha$ closer to the neat P2VP values. This behavior is observed for each silica fraction. On the other hand, the dielectric strength $\Delta\varepsilon_\alpha$ does not evolve with grafting but it decreases with the silica content even when taking into account the reduced polymer quantity (~18% reduction in PNCs with 30%v of silica at 410 K). The most important parameter for the present study is the segmental relaxation time $\tau_{max}$ as determined from the position of the maximum of the dielectric loss (α) in the raw data. It is plotted as a function of the inverse temperature in the relaxation map in Figure 2b. The curves show the classical Vogel–Fulcher–Tammann (VFT) behavior. VFT-parameters are given in SI for all PNCs in Table 1. There is a notable sequence observed in Figure 1 with surface modification at all T: the bare NPs slow-down the segmental dynamics, but grafting removes this effect, i.e., the relaxation re-accelerates back towards the one of the neat polymer. All curves but the one belonging to bare NPs in Figure 2b are practically superimposed to the neat P2VP. This superposition confirms the intuition of a saturation effect. Indeed, similar red and orange curves in Figure 1 show that grafting 0.5 or 1.1 $C_{18}$-molecules per $nm^2$ leads to similar dielectric losses around the α-peak. The fit parameters now provide a quantitative assessment of the change in average polymer dynamics.

In the inset of Figure 2b, the relaxation maps of PNCs with bare NPs are shown as a function of silica volume fraction. At high temperatures, it is found that the segmental relaxation times in PNCs tend to converge to the pure polymer timescales. As the temperature decreases, one can see a progressive slow-down of the segmental dynamics, which is more pronounced when adding more silica. Finally, the VFT-equation allows extrapolating the data to $\tau_{max}$ = 100 s to establish the correspondence with the calorimetric glass-transition temperature, $T_g$. In agreement with the relaxation maps, the highest $T_g$ (corresponding to the slowest polymer dynamics) is found for the PNCs with 30%v of bare particles: $T_g$ = 369 K vs 366 K in pure P2VP. Upon grafting, a direct return to the $T_g$ of the neat P2VP is observed, which corresponds to the removal of the slow-down effect (see SI for all $T_g$ values) and evidences the saturation effect.

**2.4 Interfacial layer dynamics in PNCs.** There is a second, more physical way the same BDS data of Figure 1 (and all other temperatures) can be analyzed. Indeed, it is plausible that the segmental dynamics is affected by the presence of the particle surfaces only up to a certain distance from the surface. This suggests partitioning of the polymer in two parts: bulk with unmodified dynamics, and an interfacial polymer layer of different segmental dynamics as described in previous works. [7, 12, 35] Such a heterogeneous IL-model is schematically depicted in Scheme 1c, where the blue part refers to the interfacial layer. It also includes the MWS and β contributions as described previously. Its application results in fits of the same high quality as those shown in Figure 1 (see SI). There is thus no possibility to distinguish which model would be more appropriate, although the ILM is physically more appealing. It also corresponds to theoretical descriptions based on the influence of the interface on local caging, and its transmission across the polymer layers up to the bulk. [37, 38] As a result, a strong gradient in timescale is present only in close vicinity of the interface, motivating the analysis based on an interlayer model. In order to display its results, we have generated a hypothetical pure interfacial contribution, $\varepsilon_{IL}^*(\omega)$, corresponding to the sum of two HN functions for the α- and β-processes. It uses the α-fit parameters obtained for each sample, the β-process being extrapolated from low-T. The IL response is compared to the neat polymer one in Figure 3 for the 20%v-PNC series. Qualitatively the same behavior is observed for 15 and 30%v of silica, and results are given in the SI. This representation highlights the modification of the segmental dynamics for the polymer part close to NP surfaces. It is slowed down in presence of bare NPs whereas it moves back to the neat polymer dynamics for the grafted NPs as shown by the shift of the α-process in Figure 1. With grafted NPs, the peak maximum is roughly at the same position, but one can still see some extra-contribution on its low-frequency side due to a broader time distribution with respect to pure P2VP. Besides, the dielectric strength of the IL $\alpha$-process is below that of the pure polymer by some 40% and it is found to be independent of grafting, consistently with our results from the first fitting approach (Scheme 1b). The observed drop in the dielectric strength of the segmental relaxation has been ascribed to reduced amplitude motions of the polymer segments in the interfacial layer.[67] In the present case, a restriction of the dipole reorientation angle by ca. 60% is sufficient to explain the dielectric response.

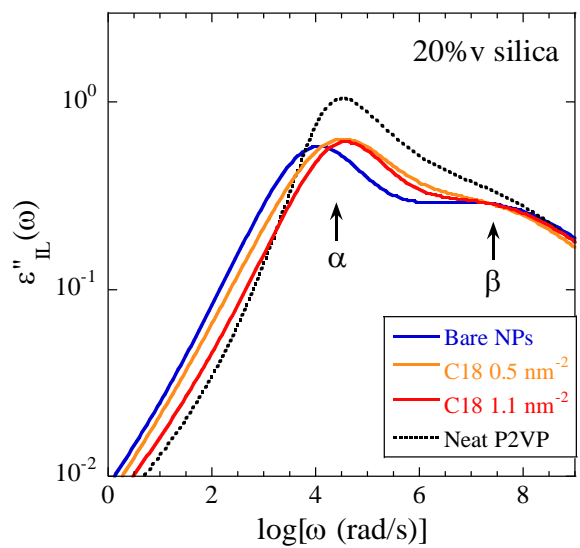

**Figure 3.** Interfacial layer contribution $\varepsilon_{IL}''(\omega)$ of the imaginary dielectric permittivity as a function of frequency (T = 423 K), including $\alpha$- and $\beta$-processes for 20%v-PNCs at different grafting densities as indicated in the legend, compared to the same processes of the neat polymer (dotted line).

The volume fraction of polymer $\Phi_{IL}$ which is part of the interfacial layer (with $\Phi_{bulk} + \Phi_{IL} = 1$) is one of the outcomes of the interfacial layer model. In Figure 4a, $\Phi_{IL}$ is plotted as a function of surface modification, for each series in silica volume fraction. Within error bars, the part of the polymer having its dynamics modified – which is the definition of the interfacial layer – is increasing with the silica content but it is remarkably independent of grafting. It has therefore been checked if the increase in volume fraction of layer corresponds to the increase in NP concentration. To see if this is the case, one needs to translate the volume fraction of the layer into a layer thickness. There are different ways to estimate this thickness. The basis is given by the available polymer volume per particle, the IL-part of which is then converted into the (identical) thickness on nanoparticles of average size. The overlap between layers of neighboring NPs at high concentrations is then tentatively estimated assuming a given spatial arrangement of the NPs, e.g., on a simple cubic lattice.[7] As the overlap tends to reduce the total quantity of layer, it results in an increase of the layer thickness at fixed $\Phi_{IL}$. The advantage of this estimation is its easy use, but it obviously fails in describing any heterogeneities in particle dispersion: in this case, the effect of particle dilution on the interparticle spacing, starting from the highest packing fraction, follows a power law (see SI), and all particles are moved away from each other correspondingly. There is thus no description of, e.g., aggregated zones and strong overlap. As we have access to representative particle positions with full polydispersity in the scattering analysis provided below, we have made use of this knowledge to obtain a better estimate of the interfacial layer thickness. The algorithm is straightforward and given in the SI. As a result, the overlap for a given particle configuration is estimated with high precision. The resulting layer thickness, i.e., the distance over which the polymer dynamics is affected away from a filler surface obtained by this method has been plotted in Figure 4b. Its average is 5±0.5 nm independent of grafting and silica content, where the error bar corresponds to the standard error (calculated from the standard deviation over 9 measurements). It is surprising to find an identical interlayer thickness for all samples, within experimental error. We believe this thickness is independent of the silica loading because it only depends on the presence of a local interface. Upon surface modification, the chain-particle interactions are weakened (towards a suppression of the slowing-down) but extend over the same distance. A physical interpretation could be that the range of the gradient is given by the transfer of cage constraints from one layer to the next, while the amplitude is a property of the interface, as suggested by molecular theories. [37, 38] A comparison with the estimated thickness assuming a cubic arrangement is given in SI, including the correction for overlapping layers at the higher concentrations. The latter leads to an average value of 4±0.5 nm, which is thus below the RMC determination that takes the heterogeneous packing structure into account. This difference illustrates the importance of having access to representative particle configurations extracted by computer simulations from the experimental measurements of particle structure by scattering.

With respect to the bare particles, the main effect of silane grafting is to re-accelerate the segmental relaxation times back to normal within a given amount of interphase, as shown by the variation of $\tau_{IL,max}$ in Figure 4c. Only the bare nanocomposites show a slow-down by a factor of 3. This is the same order of magnitude as found in previous work, [7] which also provides the ILM-average of the underlying strong (possibly double-exponential, i.e., effective over extremely short distances [37, 38]) gradient of the time scale. As soon as grafting is introduced, the value of the neat polymer is recovered. Moreover, the timescale in the interphase (as the dielectric strength) is independent of the amount of silica. This means that the interfacial dynamics is completely driven by the local polymer-NP interaction and not by the amount of NP surfaces. The ILM (Scheme 1c) thus allows to separate the impact of the presence of NP surfaces from their quantity. On the other hand, the homogeneous description (Scheme 1b) translated the increased weight of the modified layer dynamics by an artificial silica-concentration dependent shift in the average relaxation time (insert of Figure 2b). Although both approaches lead to a similar fit quality, the interlayer model seems to be more appropriate here. It not only appears to be more physical to separate the polymer into two regions, its results in terms of constant layer thickness surrounding the particles are clearly self-consistent and more appealing. In this approach, each individual layer has the same segmental relaxation time only driven by the vicinity of the silica.

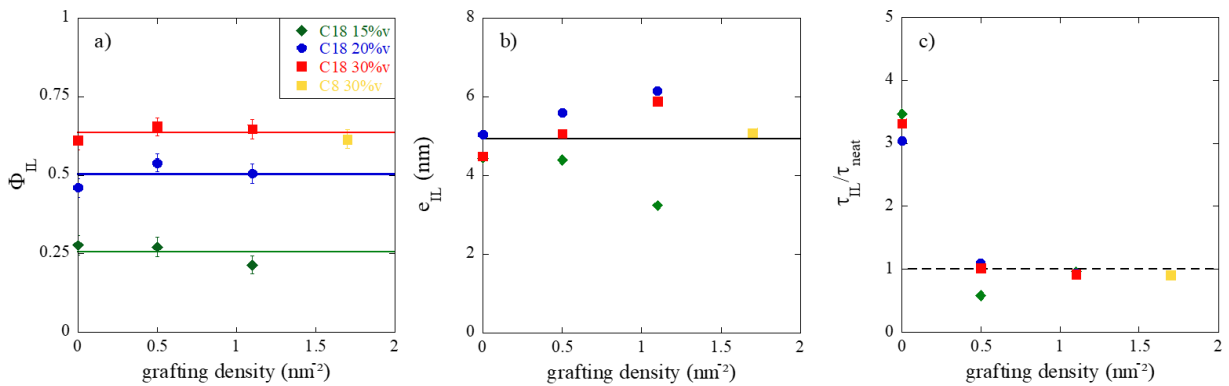

**Figure 4.** Interfacial layer volume fraction with respect to the total polymer volume **(a)** and estimated layer thickness surrounding the silica NPs as determined by the RMC NP positions **(b)**. Solid lines show the average values. **c)** Segmental relaxation time of the interfacial layer normalized to the neat P2VP timescale at 423 K (= 1.15 $T_g$). Suppression of the IL slow-down in presence of surface-modified NPs is highlighted by the dashed line. All data are plotted as a function of the silane grafting density, for different NP contents in P2VP.

Coming back to Figure 4, an additional data point has been plotted. It corresponds to a different surface modification, using a $C_8$-silane molecule instead of $C_{18}$. The resulting grafting density at a silica content of 30%v is higher in this case (1.7 nm$^{-2}$, Table 1). This result seems to confirm the saturation effect, i.e., that higher grafting densities do not further change neither the quantity of interfacial layer, nor its dynamics as observed in Figure 4.

To summarize the present study of the dynamical properties of surface-modified nanocomposites, the tuning effect of the attractive polymer-NP interactions by silane grafting has clearly been evidenced: first by applying a model of homogeneous modification of polymer dynamics following the average relaxation time of the polymer; and secondly through a precise description of the polymer interfacial dynamics taking all relaxation processes into account, simultaneously, for both the real and the imaginary part of the dielectric permittivity. The interfacial layer model provides a quantitative description of the impact of the NP surfaces on the surrounding polymer, over a distance of ca. 5 nm as obtained by coupling to the SAXS/RMC analysis given below. This impact is the same for all polymer in the vicinity of particles, only the affected volume increases with the number of NPs. Adding grafting of 0.5 molecules per nm$^2$ or more is found to screen in the same way the attractive interactions, which had slowed down the polymer dynamics in presence of bare NPs. Thus, silane grafting leads to similar interfacial layer dynamics as the neat polymer with a saturation effect. If this effect is now unambiguously demonstrated and quantitatively described, the question remains how the reduction in attractive polymer-particle interactions affect the NP-NP interactions, and thus their dispersion. This has been studied by SAXS, and analyzed by reverse Monte Carlo, in the following subsection.

**2.5 Particle dispersion and IPS distributions.** The scattered intensities for the three series at 15, 20, and 30%v of silica NPs are shown in the top row of Figure 5. The average form factor of the particles P(q) is superimposed to these data (black curve). As the intensities are normalized to the particle volume fraction, P(q) is seen to fit the data rather well in the high-q range, indicating that the silica part of the NPs (and not their surface modification) is responsible for the observed scattering pattern. At low and intermediate q, large deviations from P(q) are observed, which is obviously evidence for a non-ideal – ideality would correspond to S(q) = 1 – dispersion of particles at these high concentrations. Particle interactions are seen to be repulsive on the scale of neighboring NPs, i.e., their excluded volume leads to a low-q depression, termed the correlation hole, [52] and to a nearest neighbor peak around $q_0$. On large scales, there is possibly a low-q upturn indicating large-scale aggregation due to attractive NP-NP interactions.

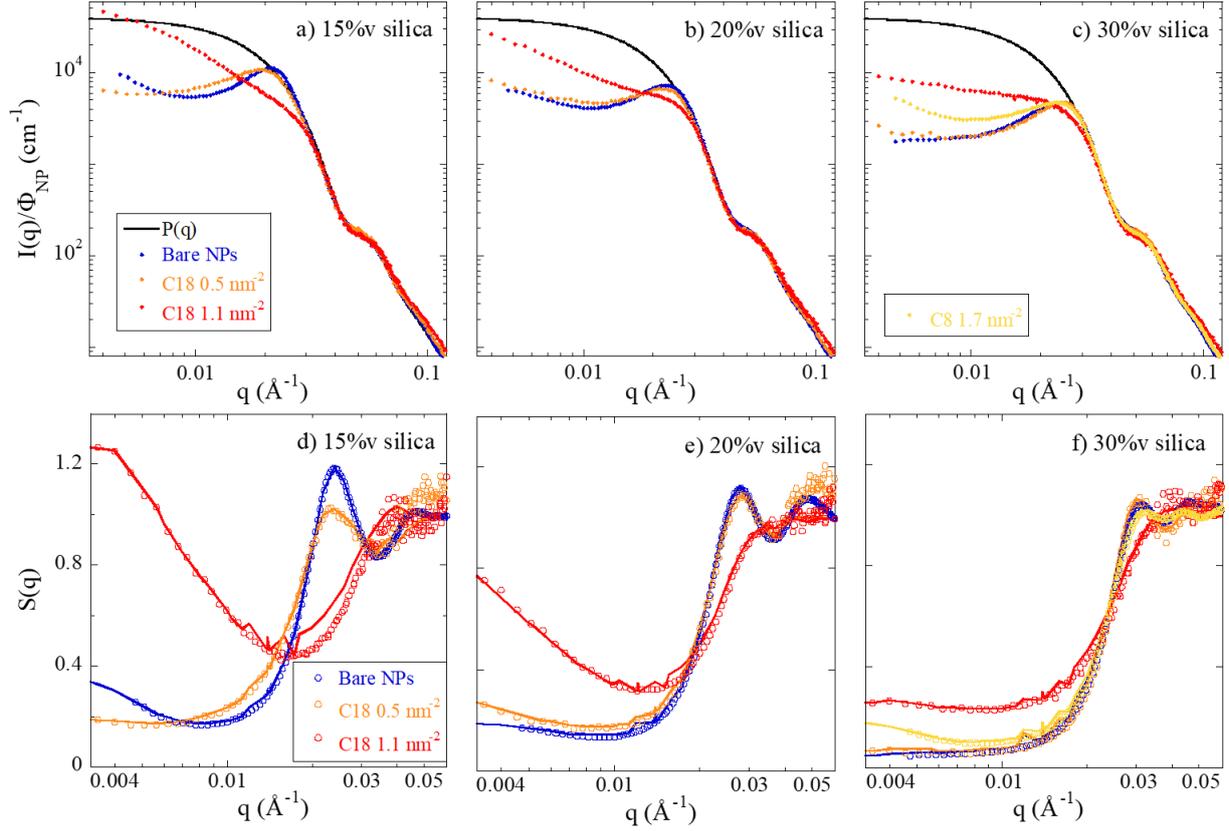

**Figure 5.** Top row: SAXS scattered intensities of P2VP-silica PNCs of different surface modifications for **(a)** 15%v-series, **(b)** 20%v, and **(c)** 30%v. The particle form factor is superimposed (black line). Bottom row: corresponding apparent structure factors with RMC fits (solid lines) for the series at **(d)** 15%v, **(e)** 20%v, and **(f)** 30%v.

As indicated above, the apparent structure factor S(q) can be obtained by dividing the experimental intensity I(q) by the average form factor P(q). The bottom row of Figure 5 shows the apparent structure factors in a log-lin representation for the same samples, together with their RMC fits. The general shape of these curves follows what can be expected from the shape of the intensities: In some cases, there is a low-q upturn, which is well visible for the highest grafting density (1.1 nm$^{-2}$). The upturn is followed by a correlation hole, and an increase towards the nearest-neighbor correlation peak between 0.02 and 0.03 Å$^{-1}$. This peak is seen to be defined best for the bare and 0.5 nm$^{-2}$ PNC samples, and somewhat ill-defined at high grafting. The same is true for the high-q structure factor in general. Similar features have been reported in nanocomposite melts by Hall et al. [68] They experimentally varied the interfacial attraction strength using polymers of different chemistry, namely poly(ethylene oxide) and polytetrahydrofuran (PTHF), the latter being less adsorbing due to a lower ability to form hydrogen bonds with the silica surface. A decrease of the magnitude of the interfacial attraction in PTHF leads to a reduced local order reflected by a low-q increase (higher compressibility) and a lower S(q) peak which is shifted towards smaller length scales. Such features are well reproduced by PRISM theory calculations, which provides a good description of polymer-mediated NP concentration fluctuations ultimately leading to depletion aggregation and microphase separation. In our case, increasing coating coverage decreases the polymer-NP effective attraction (see SI) with a qualitatively similar behavior as predicted by PRISM in terms of peak shift and low-q upturn. When comparing the three concentrations shown in Figure 5d-f, it is striking to see the narrowing of the family of intensities at higher $\Phi_{NP}$: while the three curves are largely different for the 15%v-series, they become closer and closer, with lower apparent isothermal compressibility as expressed by the low-q limits. This suggests that the effect of grafting on the dispersion is strongest at the lowest NP content, while at high $\Phi_{NP}$ the systems are already so crowded that they remain rather homogeneous and similar.

The RMC fits superimposed in Figure 5d-f correspond to the average scattering of a series of particle configurations in the simulation box. By analyzing the NP positions corresponding to the RMC fits, the surface-to-surface distances between all particles can be deduced, and their distribution function determined. The latter are shown exemplarily for the 20%v-series in Figure 6 (see SI for the others). These are the non-normalized interparticle spacing distribution functions, stating how many spheres are located in a spherical shell at a given surface-to-surface distance (with periodic boundary conditions) from a given sphere, and then averaged over all spheres present in the simulation box, and successively over the simulation run. The sum of the entries of one such function thus gives N-1, i.e., this function represents the amount of matter present around the particles in the given simulation box. Obviously, the higher the particle concentration, the more neighbors (or any couple of particles) there are, and the higher the values of this distribution function. By comparing distribution functions at different $\Phi_{NP}$ as shown in the SI, one can thus see the crowding effect.

By visual inspection of the raw IPS distributions as in Figure 6, it appears that there is a strong preference for particles to be in close contact, which depends on surface modification: the higher the grafting density, the higher the contact probability. The distribution function then decreases to a minimum of less common interparticle spacing around ca. 1 nm, before increasing again. This minimum is probably due to the exclusion of other spheres by the ones already in touch. It is noted that the pair distribution function calculated for monodisperse adhesive spheres [69] shows a similar feature in the same range. The minimum is followed by an oscillation with a small peak showing increased presence of interfaces above ca. 13 nm, which might correspond to some close but non-crystalline packing of spheres. It is also close to the polymer coil size ($2R_g$ = 10.4 nm), which may evidence the NP-free space occupied by the intercalated chains. At larger distances, finally, the average particle concentration filling the box is probed, and the number of couples of spheres in each spherical shell increases with shell volume and thus with distance, following approximately a $r^2$-law. The overall maximum at ca. $L_{box}$/2 is due to the cut-off by the finite size of the cubic simulation box, in spite of periodic boundary conditions, as particles are not counted more than once. In the raw IPS, it is difficult to see if there is any particular crowding, different from say, the structure of hard-sphere fluids without any other interaction. Moreover, even within the 20%v-series shown here, it is difficult to conclude in detail, as the experimental NP volume fractions are not exactly the same (see Table 1), and the deviations between the curves quite subtle.

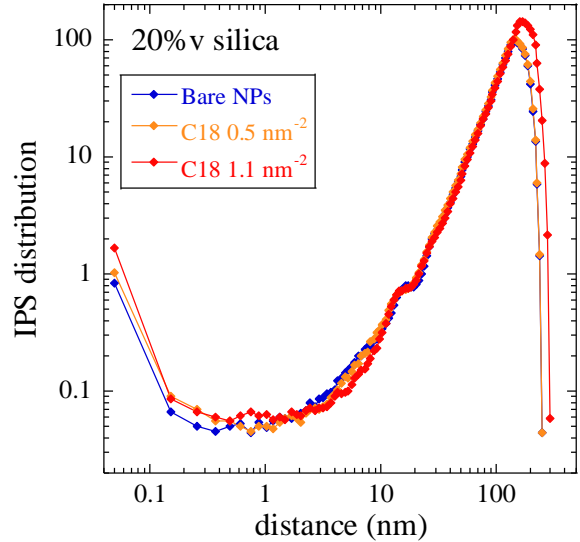

**Figure 6.** Raw interparticle surface-to-surface spacing distribution functions for the 20%v-series for bare particles, 0.5 nm$^{-2}$ and 1.1 nm$^{-2}$ grafted NPs, as indicated in the legend.

In order to proceed further with the analysis of the IPS functions, we have normalized them by the corresponding distribution function of hard-sphere fluids obeying to the same size polydispersity, and at exactly the same concentration for each sample. This normalization naturally compensates for the increase in shell volume with distance, and allows highlighting any effect of ordering induced by the polymer-mediated interactions, and by grafting. The resulting normalized IPS curves are shown in Figure 7 for the three experimental series in surface modification.

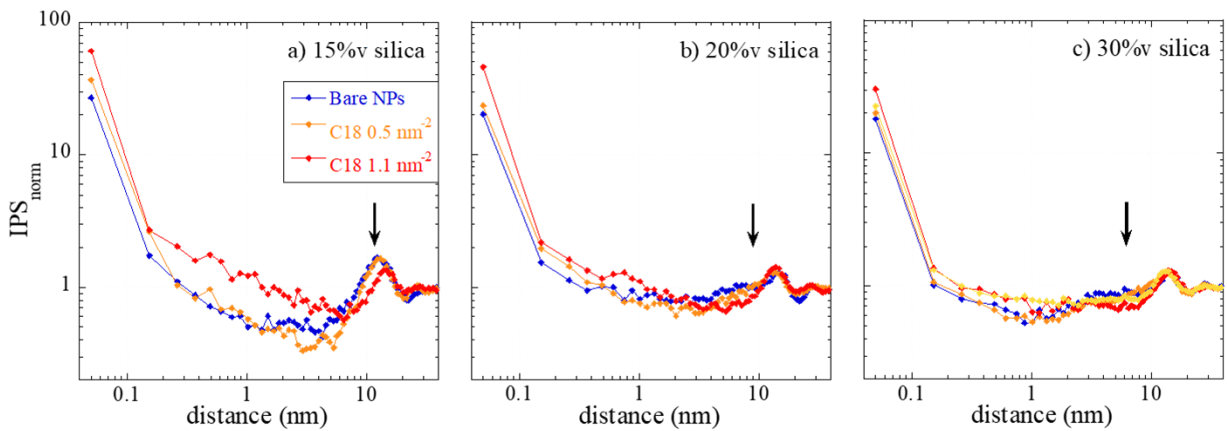

**Figure 7.** Normalized IPS for 15%v **(a)**, 20%v **(b)**, and 30%v **(c)**, each time for P2VP-silica PNCs with different surface modification as indicated in the legend. The apparent IPS deduced from the scattering peak position $q_0$ of bare and 0.5 nm$^{-2}$ grafted NPs in Figure 5 is given by an arrow (IPS = $2\pi/q_0 - 2R_{NP}$).

These IPS curves are seen to follow a similar scheme in all cases. There is a high contact value, indicating that there is a several-ten times higher direct particle contact (i.e., within the first 0.1 nm) in the nanocomposite than in the hypothetical hard-sphere fluid of identical concentration. The normed distribution then decreases to below one, indicating less neighbors at short distances (below ca. 10 nm), before displaying a peak as discussed above, and then level off to one at bigger distances, beyond a particle diameter, showing that at large scale the system is as homogeneous as a hard-sphere fluid at the same concentration. Another reassuring point of Figure 7 is that the concentration series also follows the evolution of the structure factors in Figure 5d-f: at lower concentrations both structure factors and their IPS are different, while the family of curves narrow with higher silica content, showing that the impact of grafting is strongest at low $\Phi_{NP}$.

The distribution functions in Figure 7 contain considerably more information than the commonly used $2\pi/q_0$ analysis of the position of the scattering peak. From the scattered intensities in Figure 5, $q_0$ can be used to deduce the apparent IPS of 12, 9 and 6 nm at 15, 20 and 30%v of silica, respectively, for the bare and 0.5 nm$^{-2}$ grafted NPs (arrows in Figure 7, no peak at the highest grafting). Such values are similar to the IPS calculated for a dense random structure (using $\Phi_{NP,max} = 2/\pi$).[36] They give an estimate of the strongest contribution to the center-to-center particle distance of monodisperse particles, whereas in Figure 7 the full distribution is retrieved, with account of particle polydispersity, and it works even in absence of a scattering peak. The apparent IPS are thus average values, which fail to represent the complexity of the IPS distributions and do not reflect the heterogeneity of particle packing with locally dense aggregates. Indeed, the interesting information contained in the IPS is located at small interparticle distances in Figure 7. One may mention that the features discussed above (contact, minimum, peak) are also closely related to what would be the pair-correlation function g(r-2R$_{NP}$) of an equivalent monodisperse assembly of spheres, only that here we focus on the surface-to-surface spacing between polydisperse spheres, where an average g(r) would not make sense as it does not take varying sphere sizes into account, see SI for examples and discussion.

In order to focus on local arrangement of spheres, the normalized contact values have been plotted in Figure 8a. These contact values increase systematically with grafting at all three concentrations, in agreement with the evolution of the scattered intensity and structure factors at low angles, in particular in Figure 5d, but also at higher concentrations in Figures 5e and 5f. In all cases, the low-q upturn sets in with grafting, showing that there is aggregation, and thus more close contact, which is the main quantity of interest. The increase in contact probability is found to be highest for the lowest particle volume fraction (15%v series), indicating that the biggest difference from hard-sphere fluids is obtained when there is most space for rearrangements. Due to the high density at 30%v, the dispersion is structurally closer to a (polydisperse) hard-sphere mixture at the same concentration. In other words, the particles are already quite close, and positioning some of them in close contact does not change the interparticle distances by a large amount. Therefore, any effect of grafting is necessarily less dominant at high volume fractions.

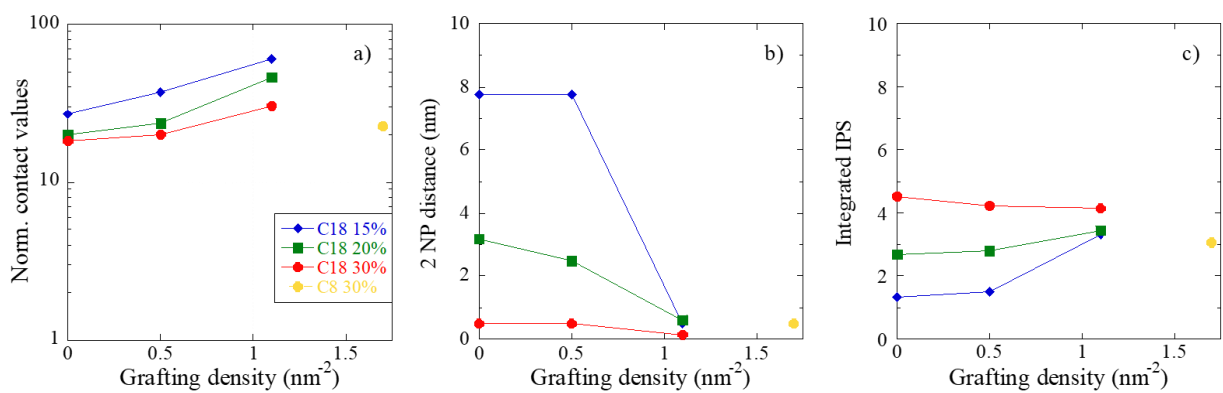

**Figure 8. (a)** Evolution of the normalized contact values with NP volume fraction (15, 20, and 30%v). **(b)** Average second-particle surface-to-surface distances determined by integration of the raw IPS up to a sum of two neighboring NPs. **(c)** Integrated raw IPS from surface to twice the silane length using 2L = 5.1 and 2.5 nm for C$_{18}$ and C$_8$, respectively. All plots are represented as a function of surface modification, for the three volume fraction series.

One may argue that the close contact excess probabilities shown in Figure 8a may be singular values covering only a thin shell (0.1 nm), and that a broader approach including a larger range of (short) distances would be more robust. We have therefore integrated the (raw, i.e., non-normalized, as shown in Figure 6) IPS distribution up to the distance when one (or two, or possibly any number smaller than N-1) neighboring spheres are encountered. This defines a distance in which on average one (or two, or any number) neighbors of a central particle are present: the smaller this distance, the more crowded the local environment. Given the many close contacts (one first particle), we chose to focus on the second-neighbor distance in order to characterize the local particle arrangements, in particular with respect to the question of polymer confinement. The result is shown in Figure 8b as a function of grafting, for the three volume fraction series. In this representation, the strongest effect of grafting is found again for the 15%v-series. The higher the grafting, the lower the second-particle distance, down to about 0.5 nm, e.g., approaching close contact for the second neighbor, too. As the NP concentration is increased, the second-particle distance decreases mechanically, because the system becomes more crowded. There is still an effect of grafting, but as stated above, its aggregating effect can only be less pronounced. The decrease in the second-particle distance is thus less strong, leading to a similar final value. At 30%v, finally, the system is already so dense that there is hardly any change induced by the surface modification.

In a last way of representing the data, it is proposed to inverse the above concept: instead of integrating up to a predetermined sum (here, distance between surfaces to the second particle) and analyze this distance, one may also integrate up to a given distance, and compare the sums. This is done in Figure 8c for the size of two $C_{18}$-silane molecules – representing the coverage of the two surfaces – in zig-zag conformation (2L = 5.1 nm). One can then read off the number of neighboring surfaces of particles possibly affected by the two grafted layers. The number of particles is found to increase with the concentration, as expected. At 30%v, there is almost no effect of grafting on this number of neighbors, whereas it increases slightly at 20%v, and more than doubles at 15%v.

The analyses shown in Figure 8 are all facets of the same evolution with concentration and surface modification. The last presentation, in Figure 8c, is the most intriguing. Intuitively, one might have expected a "buffer action" of the grafted layer, reducing the number of neighboring silica NPs allowed in close contact, at distances up to two $C_{18}$ molecules. The opposite is observed, and it can be understood in the light of the BDS results of the first part. There, it has been shown that the strong attractive interactions between the polymer and the silica NPs are effectively broken up by the surface modification, suggesting that the grafted layer pushes the polymer chains away from the NP surface, to a distance where they do not strongly interact with the surface any more. If there are then much less strongly adsorbed polymer segments on each surface, the buffer effect of the latter is reduced, and NPs can get in closer contact than the one defined by, say, two radii of gyration of the polymer chains (2$R_g$ = 10.4 nm). Apparently, the grafting density of about one per nm$^2$ of the small silane molecules is then insufficient to avoid close contact between NPs but succeeds in breaking up the close contact with the neighboring polymer chains.

Although there is strong evidence that the same process – the displacement of the chains from the NP surface by silane grafting – has a dominant, correlated impact on two completely different properties as highlighted by BDS and SAXS, a quantitative difference may be noted. While the polymer dynamics seems to saturate (Figure 1) with surface modification, i.e., very similar results are obtained for 0.5 and 1.1 nm$^{-2}$ grafting densities, this is not the case for the particle structure (Figure 5), where the evolution is more progressive (see also Figure 8). In the framework of our analysis, this implies that it is sufficient to introduce a few (0.5 nm$^{-2}$) silane molecules to screen NP-polymer interactions and to suppress the particle-surface induced slow-down in the polymer dynamics, but a higher grafting density is needed to effectively expel the polymer chains from the surface and provoke changes in NP structure. This finding may be rationalized by a partial desorption of polymer chains at intermediate grafting, where the number of monomer-NP contacts is reduced for each chain, with strong impact on the segmental dynamics because of the gain in local mobility. Only at the highest grafting, the system becomes fully non-adsorbing (see SI), and steric repulsion is strongly decreased due to the difference in size between adsorbed chain and grafted silane. The reduced steric hindrance may then be overcome by the strong interactions between the two NP surfaces. Once the chains are completely desorbed – we imagine them floating on the grafted layer – they can thus be expelled to favor direct nanoparticle contacts and attractive van der Waals interactions. Such a mechanism may also be concomitant with chain-depletion also generating interparticle attraction. The entropic depletion effect is governed by the ratio of radii ($R_{NP}$ = 9.7 nm versus $R_g$ = 5.2 nm of the polymer chain), which is not too far away from 1, suggesting a weak depletion of the chains. The sum of these effects may explain the evolution of the contact values observed in Figure 8a.

As with the BDS analysis, the additional $C_8$-modified PNC has been investigated in terms of structure. The corresponding results have been included in Figure 8, as before at higher grafting of $C_8$, at a NP concentration of 30%v. All three characteristic values of the IPS are in line with the weak evolution at this high concentration: the contact value in Figure 8a, the two-particle distance in Figure 8b, and the number of neighbors up to the grafted layer in Figure 8c, can all be seen as extrapolations (within error bars) of the $C_{18}$-data. It is concluded that $C_8$ and $C_{18}$ surface modification do not induce a qualitatively different behavior, although a more detailed study with a systematic variation of the grafting density of $C_8$ would be needed to definitely answer this issue.

## 3. CONCLUSIONS

The effect of surface modification of nanoparticles on their arrangement in space has been studied by SAXS, through an original analysis in terms of the interparticle spacing distribution function. The latter is a variant of the pair-correlation function focusing on the distance between polydisperse particle surfaces. The interparticle spacing function highlights the changes in local particle configurations, from dispersed particles to more and more aggregated ones, with increasing the grafting density. The NP configurations have been analyzed through their deviation from the ones of hard-sphere dispersions, and translated into numerical indicators, like the probability of close contact, or the number of neighbors present in a shell corresponding to the grafted layers. At the lowest NP concentration of 15%v, the effect is found to be strongest, with surface modification inducing aggregation. At higher volume fractions, where particles are close-by anyhow, the globally good NP dispersion is maintained, although some increase in close contact with grafting is evidenced. This structural evolution can be rationalized by the displacement of the polymer segments from the particle surface upon silane grafting, which disrupts the attractive polymer-particle interactions. Paradoxically, the addition of a small silane buffer thus reduces the steric buffering effect of the macromolecules on NP attraction.

The suppression of the polymer-NP interactions is found to have a strong incidence on the dynamics of the polymer layer close to the NP surface. By analyzing the BDS results and taking into account all contributing processes, the volume fraction of the interfacial polymer layer was determined. By analyzing the full particle configurations obtained by SAXS and RMC, it was possible to relate this quantity to the thickness of the interfacial layer, without artefacts induced by overlap. While the nanometric thickness of the layer of 5 nm is found to be independent of our experimental parameters ($\Phi_{NP}$ and surface modification), the segmental dynamics of the polymer chains in the interfacial layer is seen to be affected: its relaxation time is increased by the strong attractive interactions with the bare silica surfaces. Surface modification with the silane molecules tend to suppress this effect, and the dynamics of the interfacial polymer layer re-accelerates, going back towards the neat polymer dynamics with grafting. The amplitude of the gradient is thus determined by the chemistry of the interface, e.g., to which extent the interface slows down the first polymer layer by its effect on the cage, [37, 38] and/or van der Waals interactions, all modified by grafting. The spatial range of the gradient is then a consequence of transmission from monomer to monomer, and is essentially independent of the surface properties. In the ILM analysis proposed here, this amounts to observing a constant interfacial thickness, and different average dynamics.

One may wonder if there is a confinement effect triggered by grafting in this nanocomposite system. Indeed, surface modification leads to locally denser particle assemblies as shown by the IPS functions and higher contact values. This should lead to smaller and smaller polymer domains between particle surfaces and thus to a stronger geometrical confinement effect, i.e., the polymer dynamics being affected by contact with several surrounding particles. However, grafting is found to have only a weak effect on the segmental dynamics of the interfacial layer with respect to the neat polymer, with a slight broadening of the time distribution towards low frequency. This is explained by the fact that the coverage with silane molecules prevents polymer adsorption on the silica surface, and regardless of how close the surfaces are, they do not affect the dynamics anymore. Surprisingly, there are thus two competing effects of NP surface modification: it simultaneously favors polymer confinement with possibly increased impact on dynamics by changing the NP dispersion, and breaks up attractive polymer-particle interactions responsible for modification of the same dynamics. Further studies will be needed to understand if the modification of the interface can be fine-tuned, e.g., by grafting smaller molecules, as exemplified here by the tests with $C_8$-grafting, with strong relevance for both macroscopic and microscopic properties of such polymer nanocomposites.

The combination of two experimental methods in this article, BDS and SAXS, with data of both analyzed by original approaches and completed with numerical simulations, provides a self-consistent and robust picture of the important role of nanoparticle surfaces, and their chemical modification. Our combination of the methods is fruitful as it leads to the quantitative determination of the thickness of the interfacial layer, corrected for any overlap effect. Moreover, the effects of surface modification on structure and dynamics can be traced back to the same molecular origin, the location and interactions of the polymer chains with respect to surfaces. It is hoped that the striking macroscopic properties, in particular mechanical ones, of such polymer nanocomposites can thus be better understood, and possibly designed, in the future. While the interfacial layer with its dynamical slow-down (or not) has a strong impact on the deformation or flow properties of the sample, the spatial arrangement of the NPs, tunable from dispersion to aggregation, allows designing particle networks as an underlying, hard percolating structure.

## 4. MATERIALS AND METHODS

**4.1 Polymer nanocomposites.** Poly(2-vinylpyridine) with a weight-average MW of 35.9 kg.mol$^{-1}$ (polydispersity index = 1.07) was purchased from Scientific Polymer Products Inc., and used as received. The radius of gyration of the chain is 5.2 nm. The silica NPs were synthesized in ethanol by a modified Stöber method with the final NP concentration of 16 mg/mL. For the functionalization step, the NP suspension was used as is without further purification. It was characterized by SAXS and the scattered intensity revealed a log-normal size distribution ($R_0$ = 9.6 nm, σ = 17%), leading to an average NP radius of 9.74 nm.

Surface modification of the NPs was performed with *n*-octadecyldimethylmethoxysilane (CH$_3$(CH$_2$)$_{17}$Si(CH$_3$)$_2$OCH$_3$, termed $C_{18}$) and *n*-octyldimethylmethoxysilane (CH$_3$(CH$_2$)$_7$Si(CH$_3$)$_2$OCH$_3$, termed $C_8$), both from Gelest. The grafting reaction was conducted at 323 K for 3 days in ethanol. To achieve different grafting densities expressed in silane molecules per nm$^2$ of silica surface, different amounts of silanes were added to the NP suspension. For $C_{18}$, 40 mL of the suspension were mixed with 0.12 g and 0.60 g of silane resulting in NPs with the grafting densities of 0.5 and 1.1 nm$^{-2}$, respectively. For $C_8$, 50 mL of the suspension were mixed with 750 µL of silane yielding a grafting density of 1.7 nm$^{-2}$. After the reactions were completed, the surface-modified NP suspensions were dialyzed against ethanol for 3 days. The grafting densities were calculated from thermogravimetric analysis (TGA, TA instrument, Discovery, 5 K/min under air) using the weight loss between 473 and 873 K corresponding to the thermal decomposition of the grafted silanes.[70] The TGA curves are given in SI.

The polymer and (bare or surface-modified) NPs were mixed in ethanol for at least 12 hours, then filtered through a 200 nm Teflon filter. The final PNCs were formed by evaporating the solvent at room temperature followed by drying in a vacuum oven at 393 K for 2 days. All samples were hot-pressed at 423 K, and they were further annealed under vacuum at 393 K for 3 days before the SAXS and BDS measurements. The silica fractions in PNCs were obtained by TGA (20 K/min, under air) from the weight loss between 433 and 1073 K. The NP volume fractions, $\Phi_{NP}$, were determined by mass conservation using the density of neat P2VP ($\rho_{P2VP}$ = 1.19 g·cm$^{-3}$ by pycnometry)[7] and silica ($\rho_{NP}$ = 2.27 g·cm$^{-3}$ by SANS)[6]. The exact particle volume fractions are given in Table 1, where samples have been regrouped in the three series of 15, 20 and 30%v of silica.

**4.2 BDS.** BDS measurements were conducted on a broadband high-resolution dielectric spectrometer (Novocontrol Alpha) and a Quatro Cryosystem temperature controller with a stability of ± 0.1 K. The complex dielectric permittivity, $\varepsilon^*(\omega) = \varepsilon'(\omega) - i\varepsilon''(\omega)$, was measured in the frequency range from $2.10^{-2}$ to $10^7$ Hz ($\omega = 2\pi f$) using disk-shaped samples with a diameter of 20 mm and a typical thickness of 0.15 mm. The samples (without spacer) were sandwiched between two gold-plated electrodes forming a capacitor. They were first annealed for 1 h at 433 K in the BDS cryostat under nitrogen flow to ensure that both the real and imaginary parts of the permittivity became constant in the probe frequency range. Then, isothermal frequency measurements were performed from 433 K down to 368 K with an interval of 2.5 K, and from 303 K down to 233 K with an interval of 10 K to specifically follow the β-relaxation of P2VP. A measurement at the lowest measurable temperature of 103 K was performed to normalize the permittivity values. After that, the samples were measured again at 293 and 433 K to check reproducibility. The normalization procedure of PNCs is described in detail in [7] considering two-phase heterogeneous materials [24] with the high-frequency limit of the real part $\varepsilon_\infty$ = 3.05 and 3.9 for the polymer and silica, respectively. It allows getting rid of possible artifacts (mostly thickness variations) and leads to the dielectric spectra in absolute values.

**4.3 SAXS.** Small-angle X-ray measurements were performed with a wavelength λ = 1.54 Å (copper target) on two different devices. An in-house setup of the Laboratoire Charles Coulomb, "Réseau X et gamma", University of Montpellier (France) was employed using a high-brightness low-power X-ray tube, coupled with aspheric multilayer optics (GeniX$^{3D}$ from Xenocs). It delivered an ultralow divergent beam (0.5 mrad). The scattered intensities were measured by a 2D "Pilatus" pixel detector at a single sample-to-detector distance D = 1900 mm, leading to a q range from $4.7 \times 10^{-3}$ to 0.2 Å$^{-1}$. Other SAXS experiments were conducted on a SAXSLAB Ganesha with a Xenocs GeniX$^{3D}$ microfocus source at the Shared Materials Instrumentation Facility (SMIF), Duke University (North Carolina, USA). Two sample-to-detector distances were used: D = 1491 and 441 mm leading to a q range from $3.5 \times 10^{-3}$ to 0.6 Å$^{-1}$. The scattering cross-section per unit sample volume dΣ/dΩ (in cm$^{-1}$), which we term scattered intensity I(q), was obtained by using standard procedures including background subtraction and calibration.[46]

The average form factor P(q) of the nanoparticles suspended in ethanol was measured at high dilution (0.3%v), for both bare and grafted NPs, and at low concentration in polymer (ca. 2%v). In all cases, P(q) is well described by a log-normal size distribution of spheres as given above (see SI). In ethanol and in the polymer, there is thus no change in the form factor of the particles by the surface modification, because the grafted molecules do not have sufficient contrast ($\rho_{SiO2}$ = 19.49 $10^{10}$ cm$^{-2}$, $\rho_{P2VP}$ = 10.93 $10^{10}$ cm$^{-2}$, and $\rho_{C18}$ = 8.07 $10^{10}$ cm$^{-2}$). The dominant contrast is thus P2VP-silica, 8.56 $10^{10}$ cm$^{-2}$. For comparison with the PNC data, P(q) was rescaled to the appropriate contrast.

## ASSOCIATED CONTENT

**Supporting Information.** TGA data, relative attraction between P2VP and filler particles, complementary BDS and SAXS data, pair-correlation functions, and additional details on the determination of the interfacial layer thickness.

## ACKNOWLEDGMENTS

This work was supported by the U.S. Department of Energy, Office of Science, Basic Energy Sciences, Materials Sciences and Engineering Division. A.-C.G. and J.O. are thankful for support by the ANR NANODYN project, Grant ANR-14-CE22-0001-01 of the French Agence Nationale de la Recherche.

**Notes:** The authors declare that they have no competing interests.

**Table Of Contents (TOC):**

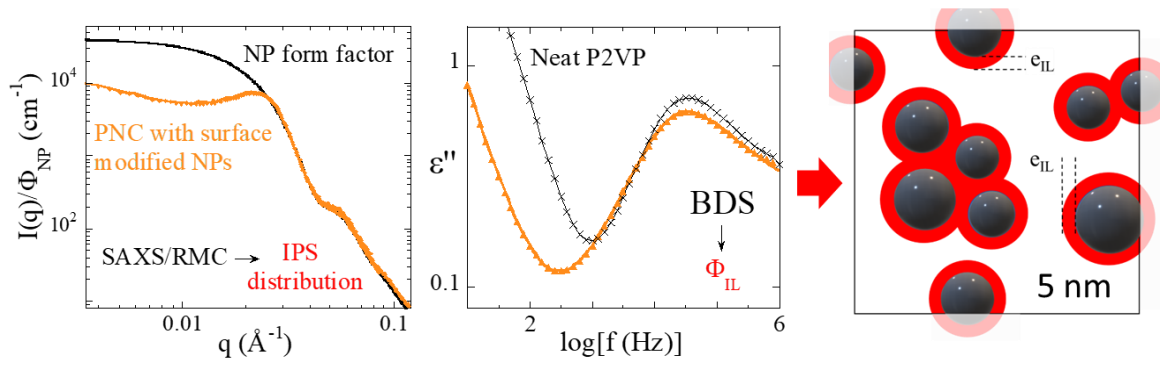